\documentclass[12pt,preprint]{aastex}

\usepackage{emulateapj5,pstricks}

\newcommand{\be}{\begin{equation}}
\newcommand{\ee}{\end{equation}}
\newcommand{\ba}{\begin{eqnarray}}
\newcommand{\ea}{\end{eqnarray}}

\shorttitle{Universal distribution for weak lensing amplification}
\shortauthors{Wang, Holz, \& Munshi}

\begin{document}

\title{A universal probability distribution function\\for
weak-lensing amplification}
\author{Yun Wang$^{1}$, Daniel E. Holz$^{2}$, and Dipak Munshi$^{3}$}
\affil{{}$^1$Department of Physics \& Astronomy,University of Oklahoma, 
Norman, OK 73019 USA. wang@mail.nhn.ou.edu\\
{}$^2$Institute for Theoretical Physics, University of California at 
Santa Barbara, Santa Barbara, CA 93106 USA\\
{}$^3$Institute of Astronomy, Madingley Road, Cambridge CB3 0HA, UK}

\begin{abstract}
We present an approximate form for the weak lensing
magnification distribution of standard candles, valid for
all cosmological models, with arbitrary matter
distributions, over all redshifts. Our results are based on
a universal probability distribution function (UPDF),
$P(\eta)$, for the reduced convergence, $\eta$.  For a given
cosmological model, the magnification probability
distribution, $P(\mu)$, at redshift $z$ is related to the
UPDF by $P(\mu)=P(\eta)/2\left|\kappa_{min}\right|$, where
$\eta=1+(\mu-1)/(2|\kappa_{min}|)$, and
$\kappa_{min}$ (the minimum convergence) can be directly
computed from the cosmological parameters ($\Omega_m$ and
$\Omega_\Lambda$).  We show that the UPDF can be well
approximated by a three-parameter stretched Gaussian
distribution, where the values of the three parameters
depend only on $\xi_\eta$, the variance of $\eta$. In short,
all possible weak lensing probability distributions can be
well approximated by a one-parameter family.  We establish
this family, normalizing to the numerical ray-shooting
results for a ${\Lambda}$CDM model by Wambsganss et
al. (1997). Each alternative cosmological model is then
described by a single function $\xi_\eta(z)$.  We find that
this method gives $P(\mu)$ in excellent agreement with
numerical ray-tracing and three-dimensional shear matrix 
calculations, and provide numerical
fits for three representative models (SCDM, $\Lambda$CDM,
and OCDM).  Our results provide an easy, accurate, and
efficient method to calculate the weak lensing magnification
distribution of standard candles, and should be useful in
the analysis of future high-redshift supernova data.

\end{abstract}

%\end{document}

%\keywords{Cosmology}


\section{Introduction}

The luminosity distance-redshift relations of cosmological
standard candles provide a powerful probe of the
cosmological parameters $H_0$, $\Omega_m$, and
$\Omega_{\Lambda}$ \citep{Garna98a,Perl99,Wang00b,Branch01}, 
as well as of the nature of the dark energy 
\citep{Garna98b,White98,Podariu00,Waga00,Maor01,Podariu01,Wang01a,Wang01b,Kujat02}. 
At present, type Ia supernovae (SNe Ia) 
are our best candidates for cosmological standard candles
\citep{Phillips93,Riess95}.  
The main systematic uncertainties of SNe Ia as cosmological 
standard candles are weak gravitational lensing 
\citep{Kantow95,frieman97,Wamb97,Holz98,HolzWald98,Wang99,Valageas00a,V00b,MJ00,Barber00,Premadi01},
and luminosity evolution \citep{Drell00,Riess99,Wang00b}.
Future SN surveys \citep[SNAP\footnote{see
http://snap.lbl.gov}]{Wang00a} could yield thousands of SNe Ia out to
redshifts of a few.  Since the effect of weak lensing
increases with redshift, the appropriate modeling of the weak
lensing of high-redshift SNe Ia will be important in the correct
interpretation of future data. In addition, with
high statistics it may be possible to directly measure
the lensing distributions, and thereby
infer properties of the dark matter~\citep{ms99,sh99}.

In general, determining the magnification distributions of standard 
candles due to weak lensing is a
laborious and time-consuming process, involving such
techniques as ray-tracing through N-body simulations or
Monte-Carlo approximations to inhomogeneous universes.
Here we present an easy, accurate, and efficient method to
calculate the weak lensing magnification distribution of standard
candles, $P(\mu)$.
Our method avails itself of a universal probability distribution function
(UPDF), $P(\eta)$, which we fit to a simple analytic form
(normalized by the cosmological N-body simulations of~\citet{Wamb97}).
All weak lensing magnification probability distributions, for all
cosmological models over all redshifts, can then be
approximated by a one-parameter family of solutions. The
underlying fundamental parameter is $\xi_\eta$, the variance of the reduced
convergence, $\eta$. To determine the magnification
PDF for a given model it is thus sufficient to determine
$\xi_\eta(z)$ for that model. We demonstrate this method with a number of
examples, and provide fitting formulae for three fiducial
cosmologies (see Table 1).

%\vspace*{-.2cm}
\begin{center}
Table 1\\
{\footnotesize{Three fiducial models}}

%{\scriptsize
{\footnotesize
\begin{tabular}{|ccccl|}
\tableline 
Model & $\Omega_m$   & $\Omega_{\Lambda}$   & $h$ & $\sigma_8$\\  
\tableline 
SCDM 		&  1.0 & 0.0 & 0.5 & 0.6 \\
$\Lambda$CDM 	&  0.3 & 0.7 & 0.7 & 0.9 \\
OCDM 		&  0.3 & 0.0 & 0.7 & 0.85 \\
\tableline
\end{tabular}
}
\end{center}

\section{Weak Lensing of Point Sources}

Due to the deflection of light by density fluctuations along the
line of sight, a source (at redshift $z_s$) will
be magnified by a factor $\mu \simeq 1+2\kappa$ (the weak lensing
limit), where the 
convergence $\kappa$ is given by \citep{Bernardeau97,Kaiser98}
\be
\kappa=\frac{3}{2}\, \Omega_m \int_0^{\chi_s} \mathrm{d}\chi\,
w(\chi, \chi_s)\, \delta(\chi),
\ee
with
\ba
\mathrm{d}\chi &=& \frac{cH_0^{-1}\, dz}{
\sqrt{\Omega_{\Lambda}+ \Omega_k
 (1+z)^2+ \Omega_m (1+z)^3}},\nonumber\\
w(\chi,\chi_s) &=& \frac{H_0^2}{c^2}\, \frac{ {\cal{D}}(\chi)\, {\cal{D}}(\chi_s-\chi)}
{{\cal{D}}(\chi_s)} \, (1+z),\nonumber\\
{\cal{D}}(\chi) &=& \frac{cH_0^{-1}}{\sqrt{|\Omega_k|}}\,
{\rm sinn}\left(\sqrt{|\Omega_k|} \, \chi\right),\nonumber
\ea
and where $\Omega_k=1-\Omega_m-\Omega_\Lambda$, and
``$\rm sinn$'' is defined as $\sinh$ if $\Omega_k>0$, and $\sin$ 
if  $\Omega_k<0$. If $\Omega_k=0$, the $\rm sinn$ and $\Omega_k$'s 
disappear. The density contrast $\delta \equiv
(\rho-\bar{\rho})/\bar{\rho}$. Since $\rho \ge 0$, 
there exists a minimum value of the convergence:
\be
\label{eq:kappamin}
\kappa_{min}= -\frac{3}{2}\, \Omega_m \int_0^{\chi_s} \mathrm{d}\chi\,
w(\chi, \chi_s).
\ee
The minimum magnification is thus given by $\mu_{min}=1+2 \kappa_{min}$. 

Now we define \citep{Valageas00a}
\be
\label{eq:eta}
\eta \equiv \frac{ \mu-\mu_{min}}{1-\mu_{min}}=
1+\frac{\kappa}{|\kappa_{min}|}
=\frac{\int_0^{\chi_s} \mathrm{d}\chi\, w(\chi, \chi_s)\, \left(\rho/\bar{\rho}\right)}
{\int_0^{\chi_s} \mathrm{d}\chi\, w(\chi, \chi_s)}.
\ee
Note that $\eta$ is the average matter density relative to the 
global mean, weighted by the gravitational lensing cross section of a 
unit mass lens along the line of sight to the source. This
is the same as the direction-dependent smoothness parameter
introduced by Wang (1999) in the weak lensing limit (Wang, in preparation).

The variance of $\eta$ is given by \citep{Valageas00a,V00b}
\be
\label{eq:xieta}
\xi_{\eta} = \int_0^{\chi_s} \mathrm{d}\chi\, \left(\frac{w}{F_s}\right)^2\,I_{\mu}(\chi),
\ee
with
\ba
F_s&=& \int _0^{\chi_s} \mathrm{d}\chi\, w(\chi, \chi_s),\nonumber\\
I_{\mu}(z)&=& \pi \int_0^{\infty} \frac{\mathrm{d}k}{k}\,\,
\frac{\Delta^2(k,z)}{k}\, W^2({\cal D}k\theta_0),
\nonumber
\ea
where $\Delta^2(k,z)= 4\pi k^3 P(k,z)$, $k$ is the wavenumber,
and $P(k,z)$ is the matter power spectrum. 
The window function $W({\cal D}k\theta_0)=2J_1({\cal D}k\theta_0)/
({\cal D}k\theta_0)$ for smoothing angle $\theta_0$. Here $J_1$ is the Bessel 
function of order 1.
Using the hierarchical ansatz to model non-linear gravitational clustering
\citep{Balian89}, Valageas (2000a,b) showed that
\be
\label{eq:Peta-V00}
P(\eta)= \int_{-i\infty}^{i\infty} \frac{\mathrm{d}y}{2\pi i\xi_{\eta}}
\, e^{ [\eta y-\phi_{\eta}(y)]/\xi_{\eta} },
\ee
where 
$\phi_{\eta}(y) \simeq \int_0^{\infty}dx\,  \left(1-e^{-xy}
\right) \, h(x)$.
The scaling function $h(x)$ can be obtained from numerical simulations
of large scale structure. For $x\ll 1$, $h(x) \propto x^{\omega-2}$
\citep{Valageas00a}, where $\omega$ is the scaling parameter.
The uncertainty in Eq.(\ref{eq:Peta-V00}) comes primarily from
the uncertainty in $\omega$.
We found that the scaling function given by Valageas (2000a)
leads to large errors in $P(\eta)$ for small $\xi_{\eta}$, making 
it less useful for calculating $P(\mu)$ at higher redshifts. 
Although $P(\mu)$ becomes increasingly
broad as source redshift increases,
$P(\eta)$ becomes increasingly {\it narrow}, since the
universe becomes more smoothly distributed at high $z$
\citep{Wang99}.

%%%%%%%%%%%%%%%%%%%%%%%%%%%%%%%%%%%%%%%%%%%%%%%%%%%%%%%%%%%%%%%%%%%%%%%%%%

\pspicture(0,0.2)(5.5,9)

\rput[tl]{0}(-0.2,10.5){\epsfxsize=8.5cm \epsfclipon
\epsffile{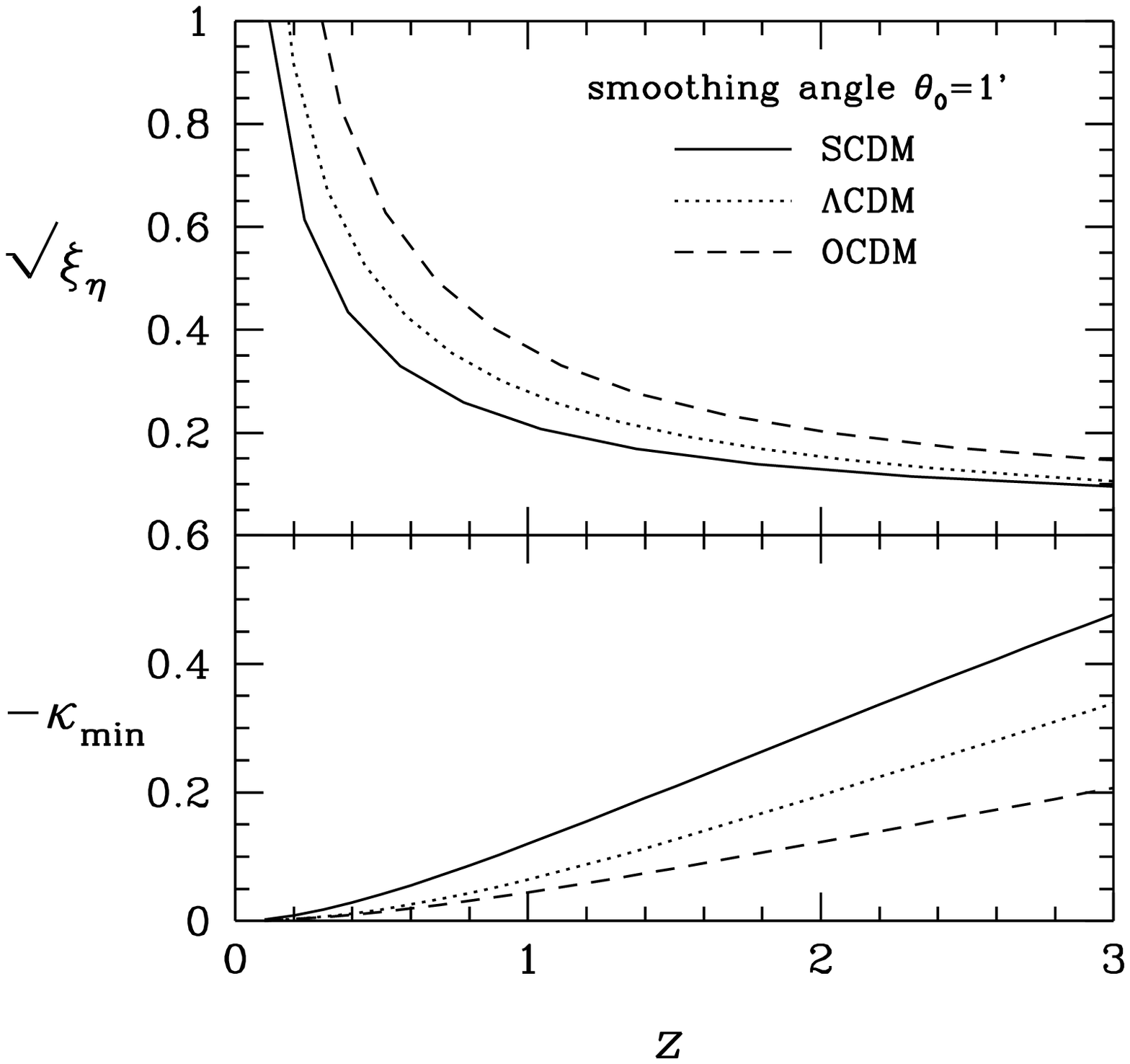}}

\rput[tl]{0}(0,2){
\begin{minipage}{8.75cm}
\small\parindent=3.5mm
{\sc Fig.}~1.---
$\sqrt{\xi_{\eta}}$ (for smoothing angle $\theta_0=1'$)
and $-\kappa_{min}$, for the three cosmological models of Table 1.
%
%\par
\end{minipage}
}
\endpspicture

%%%%%%%%%%%%%%%%%%%%%%%%%%%%%%%%%%%%%%%%%%%%%%%%%%%%%%%%%%%%%%%%%%%%%%%%%%
%\vspace*{-7.5cm}
\vspace*{-2cm}

\section{The Universal Probability Distribution Function}

Munshi \& Jain (2000) showed that $P(\eta)$ is independent of
the background geometry of the universe, as can be seen from
equation~(\ref{eq:Peta-V00}): since weak lensing contributions 
are dominated by a narrow range of the matter power spectrum, 
the scaling function $h(x)$ is independent of cosmological
parameters, and hence $P(\eta)$ has no explicit dependence
on cosmology~\citep{MJ00}. Thus the cosmological dependence of 
$P(\eta)$ enters entirely through the variance, $\xi_{\eta}$. 
We can determine the functional form of $P(\eta|\xi_{\eta})$ by 
fitting it to accurate calculations of $P(\mu)$ for {\it any} 
cosmological model. The amplification distribution, $P(\mu)$,
for arbitrary alternative cosmological models can then be found 
by computing the appropriate $\kappa_{min}$ [equation~(\ref{eq:kappamin})] 
and $\xi_{\eta}$. Utilizing $\mu =1+2|\kappa_{min}| (\eta-1)$ we find
\be
P(\mu)= \frac{P(\eta|\xi_{\eta})}{2|\kappa_{min}|}.
\label{eq:mu,P(mu)}
\ee
We call $P(\eta)$ the universal probability distribution function
(UPDF), as this one-parameter family of solutions underlies
all weak lensing magnification PDFs for {\it all} cosmologies, at 
{\it all} redshifts. We expect our results to be valid in the weak 
lensing limit, for $\kappa \la 0.2$. In particular, the PDFs derived
using our formulae are not expected to have accurate high magnification
tails.

Figure~1 shows $\sqrt{\xi_{\eta}}$ 
and $-\kappa_{min}$ computed using equations~(\ref{eq:xieta}) and
(\ref{eq:kappamin}), for the three cosmological models from Table 1.
For illustration, we give accurate fitting formulae in Table 2
for the curves in Figure 1.
\begin{center}
Table 2\\
{\footnotesize{Fitting formulae for curves in Fig.1: \\
$\sqrt{\xi_\eta}= \sum_{i=0}^3 a_i (5z)^{-i}$, 
$-\kappa_{min}= \sum_{i=0}^3 a_i (z/5)^{i}$  }}

%{\scriptsize
{\footnotesize
\begin{tabular}{|c|cccc|cccl|}
\tableline 
 & $a_0$ & $a_1$ & $a_2$ & $a_3$ & $a_0$ & $a_1$ & $a_2$ & $a_3$ \\
\tableline 
SCDM 		&  .032 & .986 & $-.452$ & .114 &  $-.025$ & .667 & .482 & $-.337$ \\
$\Lambda$CDM 	&  .021 & 1.384 & $-.642$ & .147 & $-.015$ & .280 & .766 & $-.426 $\\
OCDM 		&  .032 & 1.761 & $-.648$ & .146 & $-.004$ & .121 & .703 & $-.538 $\\
\tableline
\end{tabular}
}
\end{center}

%%%%%%%%%%%%%%%%%%%%%%%%%%%%%%%%%%%%%%%%%%%%%%%%%%%%%%%%%%%%%%%%%%%%%%%%%%

\pspicture(0,0.2)(5.5,14)

\rput[tl]{0}(-0.2,14){\epsfxsize=8.5cm \epsfclipon
\epsffile{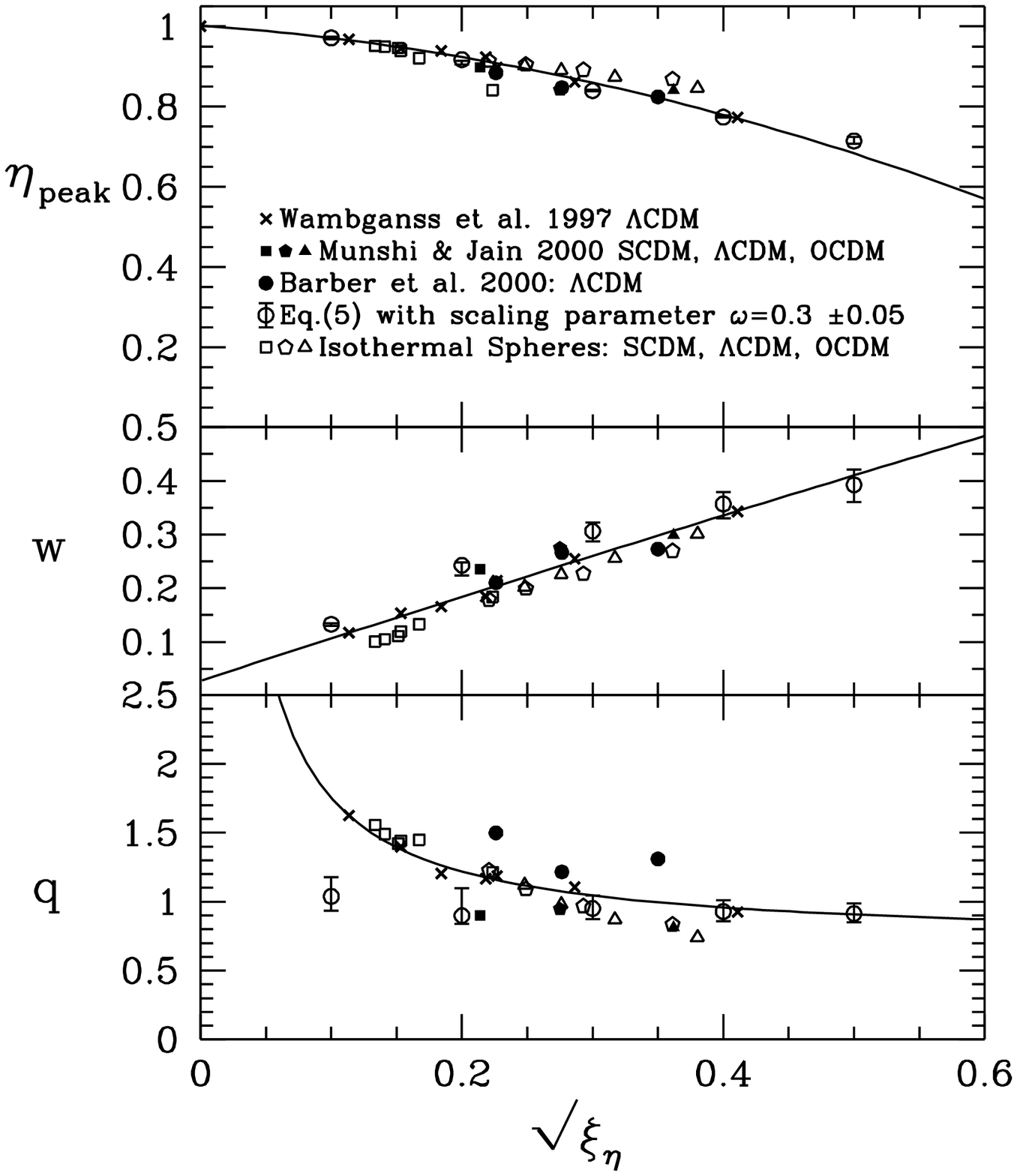}}

\rput[tl]{0}(0,5.5){
\begin{minipage}{8.75cm}
\small\parindent=3.5mm
{\sc Fig.}~2.---
The dependence of the three UPDF fitting parameters, as
functions of $\sqrt{\xi_\eta}$. Data for a variety of different
models, over a range of redshift, is shown. In addition,
best fit curves to the Wambsganss et al. points are
superposed (see equation~8).
%
%\par
\end{minipage}
}
\endpspicture
%%%%%%%%%%%%%%%%%%%%%%%%%%%%%%%%%%%%%%%%%%%%%%%%%%%%%%%%%%%%%%%%%%%%%%%%%%
\vspace*{-3cm}

We extract the UPDF, $P(\eta)$, from ray-tracing within the large
scale structure simulations of Wambsganss et al. (1997). We
then fit the UPDF to the stretched Gaussian \citep{Wang99}:
\vspace*{-0.3cm}
\be
\label{eq:P(eta)}
P(\eta|\xi_{\eta})=C_{norm}\, \exp\left[ -\left( \frac{\eta-
\eta_{peak}}
{w \,\eta^q} \right)^2 \right],
\ee
%\vspace*{-1cm}
where $C_{norm}$, ${\eta}_{peak}$, $w$, and $q$ depend
solely on
$\xi_{\eta}$ and are independent of $\eta$.
Note that although eq.(\ref{eq:P(eta)}) accurately describes 
the shape of the UPDF in the range of $\eta$ which is relevant 
to weak lensing, one must impose self-consistency by
restricting $\eta \leq \eta_{max}$, with $\eta_{max}$
chosen such that eq.(\ref{eq:P(eta)}) gives the correct 
$\xi_{\eta}$. Typically, $\eta_{max} \sim 3-7$ for $ 1\leq z
\leq 3$ in a $\Lambda$CDM model. We find that
$\langle\eta\rangle=1$ (hence $\langle\mu\rangle=1$) 
for $\eta \leq \eta_{max}$. This is because $\eta_{max}$
is sufficiently large so that the contribution of the high
$\eta$ tail to the mean is negligible.
$C_{norm}(\xi_\eta)$ is a normalization constant, chosen so
that $\int_0^{\eta_{max}}P(\eta)\,\mathrm{d}\eta=1$.  
Figure~2 shows $\eta_{peak}$, $w$, and $q$ as functions of
$\sqrt{\xi_{\eta}}$. The points denoted by crosses are extracted from the
numerical $P(\mu|z)$ by Wambsganss et al. (1997) for a
$\Lambda$CDM model with $\Omega_m=0.4$,
$\Omega_{\Lambda}=0.6$; the solid curves are ($\chi^2$
minimizing) best fits to the crosses:
%\vspace*{-2cm}
\ba
\label{eq:P(eta)par}
\eta_{peak}(\xi_{\eta}) &=&    
 1.002
-1.145    \, \left(\frac{\sqrt{\xi_{\eta}}}{5 }  \right)
-20.427  \, \left(\frac{\sqrt{\xi_{\eta}}}{5}\right)^2, \nonumber  \\
w(\xi_{\eta}) &=&
   .028 
 +3.952  \, \left(\frac{\sqrt{\xi_{\eta}}}{5 } \right) 
  -1.262    \, \left(\frac{\sqrt{\xi_{\eta}}}{5}\right)^2,\\
q(\xi_{\eta}) &=&    
    .702   
   +.509 \, \left(\frac{1}{5\sqrt{\xi_{\eta}}}   \right)
   +.008 \, \left(\frac{1}{5\sqrt{\xi_{\eta}}}\right)^2.   \nonumber
\ea   
%\vspace*{-2cm}
%\vspace*{-1cm}

The parameter ${\eta}_{peak}(\xi_\eta)$ indicates the average 
smoothness of a universe;
it increases with decreasing $\xi_\eta$ (i.e., increasing
$z$) and approaches ${\eta}_{peak}(\xi_\eta)=1$ for 
$\xi_\eta \rightarrow 0$.
The parameter $w(\xi_\eta)$ indicates the width of the distribution in
the smoothness parameter $\eta$;
it decreases with decreasing $\xi_\eta$ (i.e., increasing $z$).
The $\xi_\eta$ dependences of $\eta_{peak}(\xi_\eta)$ and 
$w(\xi_\eta)$ are as expected because as we look back to earlier times, 
lines of sight sample more of the universe,
and the universe becomes smoother on average.
The parameter $q(\xi_\eta)$ indicates the deviation of 
$P(\eta|\xi_\eta)$ from Gaussianity (which corresponds to $q=0$).

\vspace*{-0.2cm}
%\vspace*{-0.5cm}
\section{Comparison with Other Published Results}

The universal probability distribution function, $P(\eta)$,
encapsulated in eqs.~(\ref{eq:P(eta)}) and
(\ref{eq:P(eta)par}), can be used to determine the
magnification probability distribution, $P(\mu)$, for
arbitrary cosmological models at arbitrary redshifts. For
each parameter and redshift, the single free parameter
$\xi_\eta$ determines the full probability distribution.

%%%%%%%%%%%%%%%%%%%%%%%%%%%%%%%%%%%%%%%%%%%%%%%%%%%%%%%%%%%%%%%%%%%%%%%%%%

\pspicture(0,0.2)(5.5,9)

\rput[tl]{0}(-0.2,12.){\epsfxsize=8.5cm \epsfclipon
\epsffile{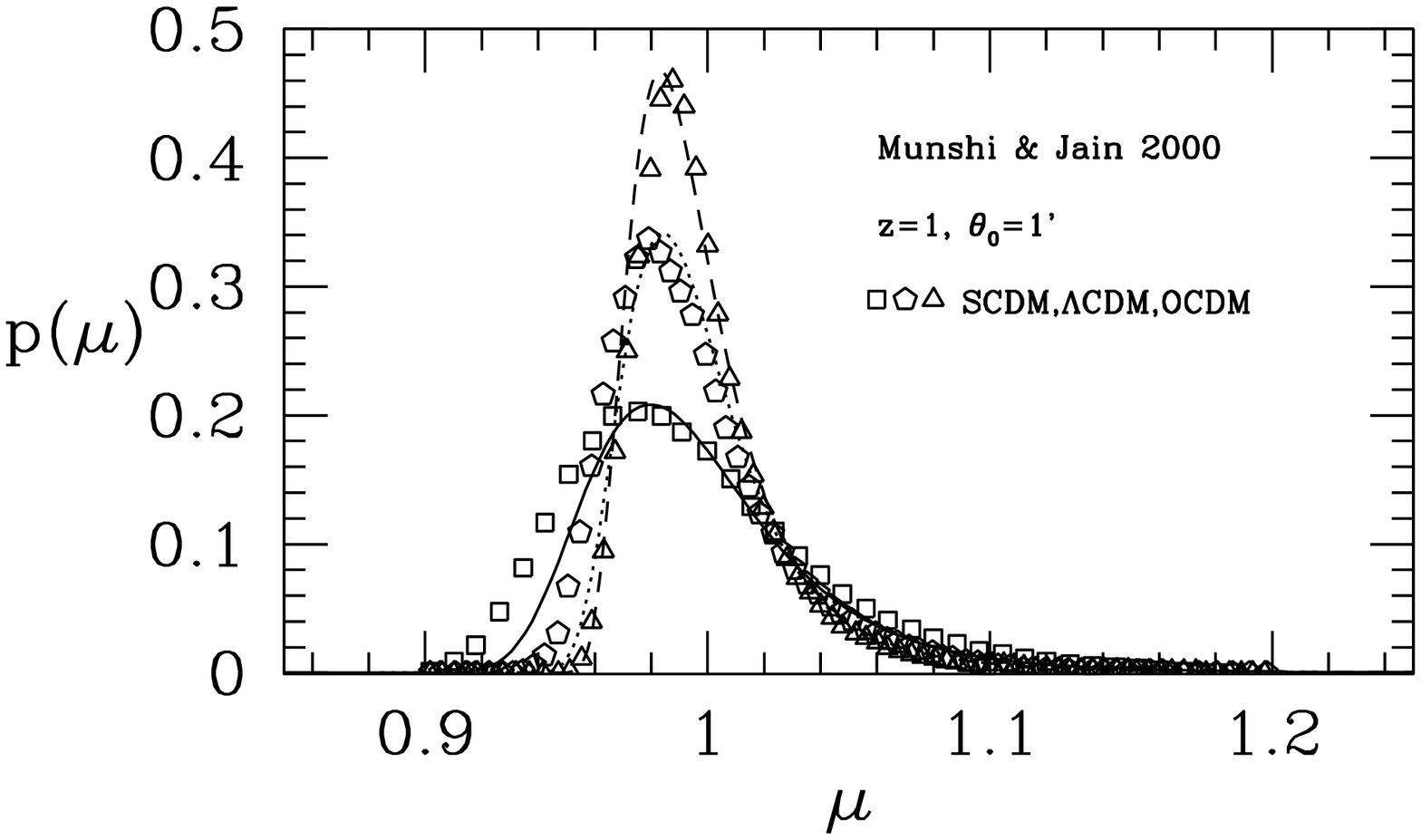}}

\rput[tl]{0}(0,4){
\begin{minipage}{8.75cm}
\small\parindent=3.5mm
{\sc Fig.}~3.--- The amplification probability distribution,
$P(\mu)$, derived from ray-tracing simulations by Munshi \&
Jain (2000) (open symbols) for smoothing angle $\theta_0=1'$,
source redshift $z_s=1$, and the three cosmological models of
Table 1,
together with $P(\mu)$ computed using our UPDF, with
$\kappa_{min}$ and $\xi_\eta$ computed using equations~(\ref{eq:kappamin}) 
and (\ref{eq:xieta}).
%
%\par
\end{minipage}
}
\endpspicture

%%%%%%%%%%%%%%%%%%%%%%%%%%%%%%%%%%%%%%%%%%%%%%%%%%%%%%%%%%%%%%%%%%%%%%%%%%
\vspace{-1.5cm}
\noindent
%%%%%%%%%%%%%%%%%%%%%%%%%%%%%%%%%%%%%%%%%%%%%%%%%%%%%%%%%%%%%%%%%%%%%%%%%%

\pspicture(0,0.2)(5.5,9)

\rput[tl]{0}(-0.2,12.){\epsfxsize=8.5cm \epsfclipon
\epsffile{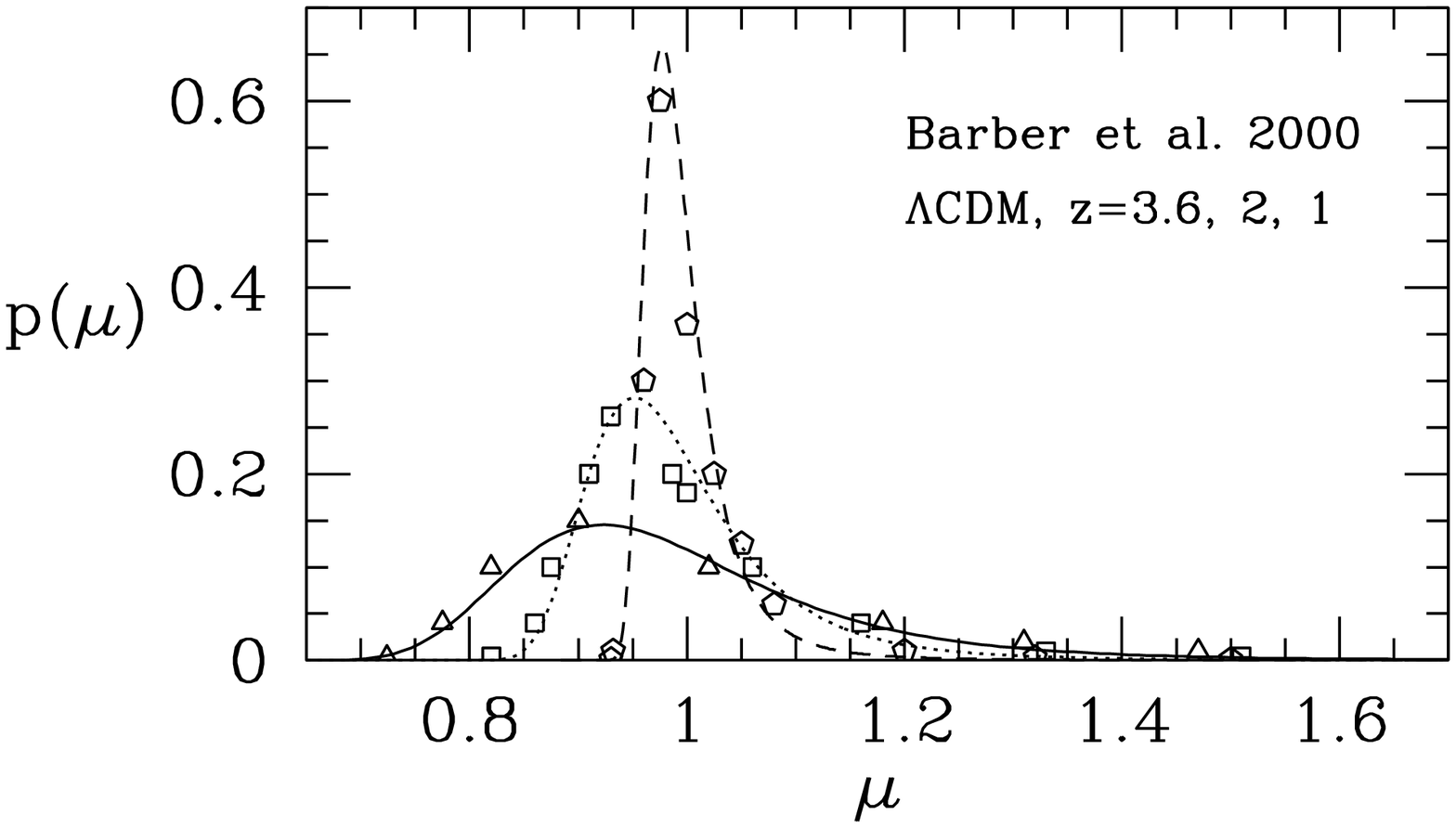}}

\rput[tl]{0}(0,4){
\begin{minipage}{8.75cm}
\small\parindent=3.5mm
{\sc Fig.}~4.---
The $P(\mu)$ from three-dimensional shear matrix calculations of
N-body simulations by Barber et al.
(2000) (circles) for a $\Lambda$CDM model with $\Omega_m=0.3$, 
$\Omega_{\Lambda}=0.7$ at source redshifts $z_s=3.6$, 2, 1 (peaking
from left to right), together 
with $P(\mu)$ computed using our UPDF,
with $\kappa_{min}$ computed using equation~(\ref{eq:kappamin}) and
$\xi_\eta$ inferred from Table 4 of Barber et al. (2000).
%
%\par
\end{minipage}
}
\endpspicture

%%%%%%%%%%%%%%%%%%%%%%%%%%%%%%%%%%%%%%%%%%%%%%%%%%%%%%%%%%%%%%%%%%%%%%%%%%
%\vspace{-1.5cm}
\vspace{-1.cm}
\noindent

Figure~3 shows the $P(\mu)$ from ray-tracing simulations by
Munshi \& Jain (2000) for smoothing angle
$\theta_0=1'$, source redshift $z_s=1$, and three
cosmological models from Table 1, together with $P(\mu)$ computed using
our UPDF for the $\kappa_{min}$ and $\xi_\eta$ computed using
equations~(\ref{eq:kappamin}) and (\ref{eq:xieta}).

Figure~4 shows the $P(\mu)$ from three-dimensional shear matrix calculations of
N-body simulations by Barber et al.
(2000) for a $\Lambda$CDM model with $\Omega_m=0.3$, 
$\Omega_{\Lambda}=0.7$ at source redshifts $z_s=1$, 2, 3.6, together 
with $P(\mu)$ computed using our UPDF,
with $\kappa_{min}$ computed using equation~(\ref{eq:kappamin}) and
$\xi_\eta$ inferred from Table 4 of Barber et al. (2000)
[equation~(\ref{eq:xieta}) was not used to compute $\xi_\eta$
due to our lack of knowledge 
of the smoothing angle $\theta_0$ that corresponds to their results].

Our UPDF gives $P(\mu)$ in excellent agreement with
N-body calculations. To make this more apparent, we have
extracted $P(\eta)$ from the $P(\mu)$ obtained via
N-body calculations by Munshi \& Jain (2000) and Barber et
al. (2000), and fitted them to the functional form of
equation~(\ref{eq:P(eta)}). The resultant coefficients are
plotted in Figure~2. There is very good agreement in the
peak location $\eta_{peak}(\xi_{\eta})$ and width indicator
$w(\xi_\eta)$, but larger scatter in the non-Gaussianity
indicator $q(\xi_{\eta})$ extracted from Munshi \& Jain
(2000) and Barber et al. (2000).  The latter may arise
partly due to the fact that in both cases we poorly resolve 
the non-Gaussian tails, which are crucial to determining accurate 
values of $q$. 
In addition, the weak lensing condition breaks down for the high 
$\mu$ tails, which could be significant for small $\xi_{\eta}$.
Also plotted in Figure~2
are the coefficients extracted from fitting the analytically
computed $P(\eta)$ [see Eq.(\ref{eq:Peta-V00})], following 
Munshi \& Jain (2000), for the
scaling parameter $\omega=0.3\pm 0.05$. These $P(\eta)$ have
not been tested for $z>1$ 
(i.e., for small $\xi_{\eta}$),
although the deviations are expected to be small, since
$P(\eta)$ peaks close to $\eta=1$ at small $\xi_{\eta}$ [see
equation~(\ref{eq:P(eta)})].

Figure~2 also shows the $P(\eta)$ coefficients extracted from ray-tracing
of randomly placed singular isothermal spheres (SIS), following the
prescription of Holz \& Wald (1998), for the three
cosmological models of Table 1; these are in good
agreement with the fitted coefficients from Wambsganss et al. (1997).
We find that improved statistics leads to better agreement 
between the $q(\xi_\eta)$ from our ray-tracing of randomly 
placed mass distributions and that from Wambsganss et al. (1997),
while having much less impact on $\eta_{peak}(\xi_{\eta})$ and $w(\xi_{\eta})$.
This is as expected, since improved statistics fills out the non-Gaussian
tails of the $P(\mu)$. 

%\vspace{-1.5cm}
\section{Summary and Discussion}

We have derived a simple and accurate method to compute the 
weak lensing magnification distribution, $P(\mu)$, for 
standard candles placed at any redshift in arbitrary cosmological 
models. We use a universal probability 
distribution function (UPDF), $P(\eta| \xi_{\eta})$, which
is independent of cosmological model; the dependence on
cosmology entering only through the variance, $\xi_{\eta}$, of
the reduced convergence, $\eta$.
The UPDF is fit accurately by a 3-parameter stretched Gaussian 
distribution [eq.~(\ref{eq:P(eta)})]. We
give polynomial fitting formulae
[eq.~(\ref{eq:P(eta)par})] for the three parameters
$\eta_{peak}(\xi_{\eta})$ (average smoothness),
$w(\xi_{\eta})$ (smoothness variation), and $q(\xi_{\eta})$
(non-Gaussianity), which we normalize to the N-body
simulations of Wambsganss et al. (1997).  The magnification
PDF, $P(\mu, z)$, can then be determined from the UPDF
using~equation~(\ref{eq:mu,P(mu)}).
We expect our results to be valid in the weak lensing limit,
for $\kappa \la 0.2$. The extension of our method to high
magnifications will be presented elsewhere.

To test the robustness of this method, we have compared our
results against three alternate independent methods (see
Fig.~2).  We find excellent agreement with: 
(1) the N-body calculations of Munshi \& Jain (2000) and
Barber et al. (2000), with some scatter in the
non-Gaussianity indicator $q(\xi_{\eta})$, which 
is consistent with the limited statistics at low $z$ 
(i.e., large $\xi_{\eta}$) of these N-body results
and the breaking down of the weak lensing condition at high 
$\mu$, 
(2) the analytical calculation [see equation~(\ref{eq:Peta-V00})]
following Munshi \&
Jain (2000), where the latter has been verified by ray tracing 
experiments, and (3) the ray-shooting of randomly placed
SIS mass distributions, following the prescription of Holz
\& Wald (1998).

We expect these simple, universal forms for the weak lensing
distribution to be useful in addressing high redshift
data, and in particular, in the analysis of results from future
supernova surveys.

\acknowledgements

We thank Joachim Wambsganss for providing us with the
results of his N-body determinations of $P(\mu)$, 
Bhuvnesh Jain, Ron Kantowski, Alex Kim and
Eric Linder for useful discussions,
and the referee for helpful comments and suggestions.
This work was supported in part by 
NSF CAREER grant AST-0094335 (YW),
NSF grant PHY99-07949 to the ITP (DEH), 
and a PPARC grant at IOA (DM).

\end{document}